\documentclass{ws-procs975x65}
\begin{document}

\long \def \blockcomment #1\endcomment{}
 
\title{The lattice and quantized Yang-Mills theory}
  
\author{Michael Creutz}

\address{Physics Department, Brookhaven National Laboratory\\
  Upton, NY 11973, USA\\
 E-mail: creutz@bnl.gov}

\begin{abstract}
Quantized Yang-Mills fields lie at the heart of our understanding of
the strong nuclear force. To understand the theory at low energies, we
must work in the strong coupling regime. The primary technique for
this is the lattice. While basically an ultraviolet regulator, the
lattice avoids the use of a perturbative expansion. I discuss the
historical circumstances that drove us to this approach, which has had
immense success, convincingly demonstrating quark confinement and
obtaining crucial properties of the strong interactions from first
principles.
\end{abstract}


\bodymatter

\section{Introduction}

As we have been hearing throughout this meeting, the Yang-Mills theory
\cite{Yang:1954ek} was developed in an attempt to generalize the gauge
symmetry of electromagnetism to the non-Abelian SU(2) symmetry of
isospin.  Remarkably, this simple idea has developed into a core
ingredient of all modern theories of elementary particles.  With the
particular application to the strong interactions, quarks interact by
exchanging non-Abelian gauge gluons.  This gives rise to some rather
unique issues.  In particular, asymptotic freedom and dimensional
transmutation imply that low energy physics is controlled by large
effective coupling constants.  Long distance phenomena, such as chiral
symmetry breaking and quark confinement, lie outside the realm of
accessibility to the traditional Feynman diagram approach.  This has
driven theorists to new approaches, amongst which the lattice has
proven the most successful.

In this talk I will present an overview of what motivated the lattice
approach and how it grew to become the dominant technique to study
non-perturbative effects in quantum field theory.  Much of this
presentation is adapted from a previous review\cite{Creutz:2004eq}.
Along the way we will see that there are both practical and
fundamental issues with the lattice method.  On the practical side,
quantitative computer calculations are now routine for
non-perturbative effects in the strong interactions.  On the more
conceptual side, the lattice gives deep insights into the workings of
relativistic field theory.

\section {Before the lattice}

I begin with the situation in particle physics in the late 60's, when
I was a graduate student.  Quantum-electrodynamics was the model field
theory, with immense success.  While hard calculations remained, and
indeed still remain, the feeling was that this theory was understood.
Some subtle conceptual issues do still remain, such as the likely
breakdown of the perturbative expansion at ultra high energies.

These were the years when the ``eight-fold way'' for describing
multiplets of particles had gained widespread acceptance
\cite{GellMann:1964xy}.  The idea of ``quarks'' was around, but with
considerable caution about assigning them any physical reality; were
they nothing but a useful mathematical construct?  A few insightful
theorists were working on the weak interactions, and the basic
electroweak unification was emerging
\cite{Weinberg:1967tq,Salam:1968rm}.  The SLAC experiments were
observing substantial inelastic electron-proton scattering at large
angles.  This was quickly interpreted as evidence for substructure,
and the idea of ``partons'' became popular.  While there were
speculations on connections between quarks and partons, people tended
to be rather cautious about pushing this too hard.

A crucial feature of the time was the failure of extension of quantum
electrodynamics to a meson-nucleon field theory.  The pion-nucleon
analog of the electromagnetic coupling had a value about 15, in
comparison with the 1/137 of QED.  This meant that higher order
corrections to perturbative processes were substantially larger than
the initial calculations.  There was no known small parameter in which
to expand.

In frustration over this situation, much of the particle theory
community abandoned traditional quantum field theoretical methods and
explored the possibility that particle interactions might be
completely determined by fundamental postulates such as analyticity
and unitarity.  This ``S-matrix'' approach raised the deep question of
just ``what is elementary?''  A delta baryon might be regarded as a
combination of a proton and a pion, but it would be just as correct to
regard the proton as a bound state of a pion with a delta.  All
particles were to be thought of as bound together by exchanging
themselves \cite{Chew:1965fda}.  These ``dual'' views of the basic
objects of the theory have evolved into many of the ideas of string
theory.

\section{ The birth of QCD}

In the early 1970's, partons were increasingly identified with quarks.
This shift was pushed by two dramatic theoretical accomplishments.
First was the proof of renormalizability for Yang-Mill's
theories\cite{'tHooft:1972fi}, giving confidence that these elegant
mathematical structures\cite{Yang:1954ek} might indeed have something
to do with reality.  Second was the discovery of asymptotic freedom,
the fact that interactions in Yang-Mills theories become weaker at
short distances\cite{Politzer:1973fx,Gross:1973id}.  Indeed, this was
quickly connected with the point-like structures hinted at in the SLAC
experiments.  Out of these ideas evolved QCD, the theory of quark
confining dynamics.

The viability of this picture depends upon the concept of
``confinement.''  While there exists strong evidence for quark
substructure, no free quarks have ever been observed.  This is
particularly puzzling given the nearly free nature of their
interactions inside the nucleon.  Indeed, the question of ``what is
elementary?''reappears.  Are the fundamental objects the physical
particles we see in the laboratory or are they these postulated quarks
and gluons?

\begin{figure}
\centering
\includegraphics[width=.7\hsize]{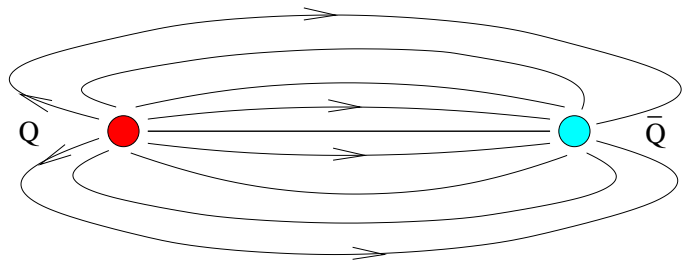}
\caption {A tube of gluonic flux connects quarks and anti-quarks.  The
strength of this string is 14 tons.}
\label{fluxtube}
\end{figure}

Struggling with this paradox led to the now standard flux-tube picture
of confinement.  The eight gluons are analogues of photons except that
they carry ``charge'' with respect to each other.  Gluons would
presumably be massless like the photon were it not for confinement.
But a massless charged particle would be a rather peculiar object.
Indeed, what happens to the self energy in the electric fields around
a gluon?  Such questions naturally lead to a conjectured instability
of the {\ae}ther that removes zero mass gluons from the physical
spectrum.  This should be done in a way that does not violate Gauss's
law.  Note that a Coulombic $1/r^2$ field is a solution of the
equations of a massless field, not a massive one.  Without massless
particles in the spectrum, such a spreading of the gluonic flux is not
allowed since it cannot satisfy the appropriate equations in the weak
field limit.  According to Gauss's law, the field lines emanating from
a quark cannot simply end.  Instead of spreading in an inverse square
manner, the gluo-electric flux lines cluster together, forming a tube
emanating from the quark and ultimately ending on an anti-quark as
sketched in Fig.~{\ref{fluxtube}}.  This structure should be regarded
as a real physical object, which grows in length as the quark and
anti-quark are pulled apart.  The resulting force is constant at long
distance, and is measured via the spectrum of high angular-momentum
states, organized into the famous ``Regge trajectories.''  In physical
units, the flux tube pulls with a tension of about 14 tons.

In essence, the reason a quark cannot be isolated is similar to the
fact that a piece of string cannot have just one end.  Of course a
piece of string can't have three ends either. This is resolved by the
underlying $SU(3)$ group theory, wherein three fundamental charges can
form a neutral object.  It is important to emphasize that the
confinement phenomenon cannot be seen in perturbation theory; when the
coupling is turned off, the spectrum becomes free quarks and gluons,
dramatically different than the pions and protons of the interacting
theory.

\section{ The 70's revolution}

The discoveries related to the Yang-Mills theory were just the
beginning of a revolutionary period in particle physics.  Perhaps the
most dramatic event was the discovery of the $J/\psi$
particle\cite{Eichten:1979ms}.  The interpretation of this object and
its partners as bound states of heavy quarks provided what might be
regarded as the hydrogen atom of QCD.  The idea of quarks became
inescapable; field theory was reborn.  The $SU(3)$ non-Abelian gauge
theory of the strong interactions was combined with the recently
developed electroweak theory to become the durable ``standard model.''

This same period also witnessed several remarkable events on a more
theoretical front.  Non-linear effects in classical field theories
were shown to have deep consequences for their quantum counterparts.
Classical ``lumps'' represented a new way to get particles in a
quantum field theory\cite{Coleman:1975qj}.  Much of the progress here
was in two dimensions, where techniques such as ``bosonization''
showed equivalences between theories of drastically different
appearance.  A boson in one approach might appear as a bound state of
fermions in another, but in terms of the respective Lagrangian
approaches, they were equally fundamental.  Again, we were faced with
the question of ``what is elementary?''

The interest in classical solutions quickly led to the discovery
\cite{Belavin:1975fg} of ``pseudo-particles'' or ``instantons,''
solutions of the four dimensional Yang-Mills theory in Euclidean space
time.  These turned out to be intimately related to the famous
anomalies in current algebra, and gave a simple mechanism to generate
the masses of such particles as the $\eta^\prime$
\cite{'tHooft:1976fv}.  These effects are inherently non-perturbative,
having an explicit exponential dependence in the inverse coupling.

This slew of discoveries had deep implications: field theory had many
aspects that could not be seen via the traditional analysis of Feynman
diagrams.  This has crucial consequences for practical calculations.
Field theory is notorious for divergences requiring regularization.
The bare mass and charge are divergent quantities.  They are not
physical observables.  For practical calculations, a ``regulator'' is
required to tame the divergences, and when physical quantities are
related to each other, any regulator dependence should drop out.

The need for controlling infinities had been known since the early
days of QED.  But all regulators in common use were based on Feynman
diagrams; one would calculate until a divergent diagram appeared, and
that diagram was then cut off.  Numerous schemes were devised for this
purpose, ranging from the Pauli-Villars approach \cite{Pauli:1949zm}
to the forest formulae \cite{Gomes:1974cr} to dimensional
regularization \cite{'tHooft:1973mm}.  But with the increasing
realization that non-perturbative phenomena were crucial, it was
becoming clear that we needed a ``non-perturbative'' regulator,
independent of Feynman diagrams.

\section{The lattice}

The necessary tool appeared with Wilson's lattice theory.  He
originally presented this as an example of a model exhibiting
confinement.  The strong coupling expansion has a non-zero radius of
convergence, allowing a rigorous demonstration of confinement, albeit
in an unphysical limit.  The resulting spectrum has exactly the
desired properties; only gauge singlet bound states of quarks and
gluons can propagate.

This was not the first time that the basic structure of lattice gauge
theory had been written down.  A few years earlier,
Wegner\cite{Wegner:1984qt} presented a $Z_2$ lattice gauge model as an
example of a system possessing a phase transition but not exhibiting
any local order parameter.  In his thesis, Jan Smit\cite{smitthesis}
described using a lattice to formulate gauge theories outside of
perturbation theory.  Very quickly after Wilson's suggestion, Balian,
Drouffe, and Itzykson \cite{Balian:1974ts,Balian:1974ir,Balian:1974xw}
explored an amazingly wide variety of aspects of these models.

To reiterate, the primary role of the lattice is to provide a
non-perturbative regulator.  Space-time is not really meant to be a
crystal; the lattice is a mathematical trick.  It provides a minimum
wavelength through the lattice spacing $a$, {\it i.e.} a maximum
momentum of $\pi/a$.  Path summations become well defined ordinary
integrals.  By avoiding the convergence difficulties of perturbation
theory, the lattice provides a route towards a rigorous definition of
a quantum field theory as a limiting process.

This approach had a marvelous side effect.  After discreetly making
the system discrete, the lattice system becomes sufficiently well
defined to be placed on a computer\cite{Creutz:1980zw}.  This was fairly
straightforward, and came at the same time that computers were growing
rapidly in power.  Indeed, numerical simulations and computer
capabilities have continued to grow together, making these efforts the
mainstay of modern lattice gauge theory.

\section{Gauge fields and phases}

As formulated by Wilson, the lattice cutoff is remarkable in remaining
true to many of the underlying concepts of a gauge theory.  At the
most simplistic level, a Yang-Mills theory is simply electrodynamics
embellished with isospin symmetry.  By working directly with elements
of the gauge group, this is inherent in lattice gauge theory from the
start.

At another level, a gauge theory is a theory of phases acquired by a
particle as it passes through space time.  Using group elements on
links directly gives this connection, with the phase associated with
some world-line being the product of these elements along the path in
question.  Ror the Yang-Mills theory, the concept of ``phase'' becomes
a rotation in the internal symmetry group.

At a still deeper level, a gauge theory is a theory with a local
symmetry.  With the Wilson action being formulated in terms of
products of group elements around closed loops, this symmetry remains
exact even with the regulator in place.

In perturbative discussions, the local symmetry forces one to adopt a
gauge-fixing prescription to remove a formal infinity on integrating
over gauges.  The lattice formulation, in contrast, uses a compact
representation for the group elements, making the integration over all
gauges finite.  For gauge invariant observables, no gauge fixing is
required.  While gauge fixing can still be done, and must be
introduced to study such conventional gauge-variant quantities such as
gluon or quark propagators.

One property of continuum gauge theory that the lattice approach
violates involves transformations under Lorentz transformations.  In a
continuum theory the basic vector potential can change under a gauge
transformation when transforming between frames\cite {Weinberg:1964cn,
  Weinberg:1964ev, Weinberg:1969di}.  The lattice, of course, breaks
Lorentz invariance, and thus this concept loses meaning until the
continuum limit is taken.
 
\section{The Wilson action}

\begin{figure}
\centerline{
\includegraphics[width=.45\hsize]{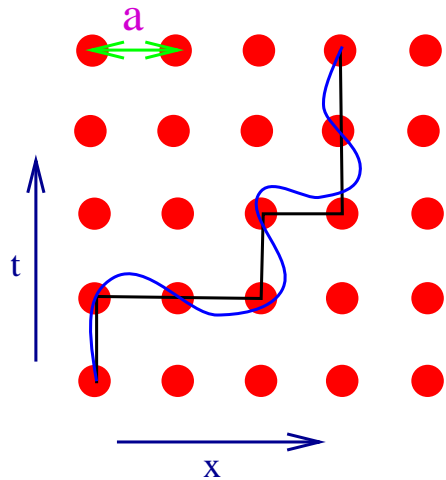}
}
\caption{In lattice gauge theory the world-line describing the motion
of a quark through space-time is approximated by a sequence of
discrete hops.  On each of these hops the quark wave function picks up
a ``phase'' described by the gauge fields.  For the strong
interactions, this phase is a unitary matrix in the group $SU(3)$.
}
\label{worldline}
\end{figure}

The concept of gauge fields representing path dependent phases leads
directly to the conventional lattice formulation.  We approximate a
general quark world-line by a set of hoppings lying along lattice
bonds, as sketched in Fig. \ref {worldline}.  We then introduce the
gauge field as group valued matrices on these bonds.  Thus the gauge
fields form a set of $SU(3)$ matrices, one such associated with every
nearest neighbor bond on our four-dimensional hyper-cubic lattice.

\begin{figure}
\centerline{
\includegraphics[width=.3\hsize]{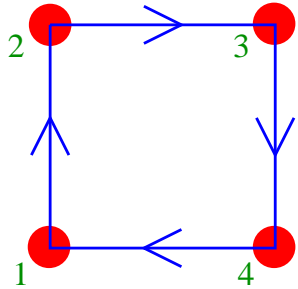}
}
\caption{Analogous to Stoke's law, the flux through an elementary
  square of the lattice is found from the product of gauge matrices
  around that square.  The dynamics is determined by adding the real
  part of the trace of this product over all elementary squares.  This
  ``action'' is inserted into a ``path integral.''  The resulting
  construction is formally equivalent to a partition function for a
  system of ``spins'' existing in the group $SU(3)$.  }
\label{plaquette}
\end{figure}

In terms of these matrices, the gauge field dynamics takes a simple
natural form.  In analogy with regarding electromagnetic flux as the
generalized curl of the vector potential, we are led to identify the
flux through an elementary square, or ``plaquette,'' on the lattice
with the phase factor obtained on running around that plaquette; see
Fig. \ref {plaquette}.  Spatial plaquettes represent the ``magnetic''
effects and plaquettes with one time-like direction give the
``electric'' fields.  This motivates the conventional ``action'' used
for the gauge fields as a sum over all the elementary squares of the
lattice.  Around each square we multiply the phases and to get a real
number we take the real part of the trace
\begin{equation}
S_g=\sum_p {\rm Re\ Tr} \prod_{l\in p} U_l \sim \int d^4x\ E^2+B^2.
\end{equation}  
Here the fundamental squares are denoted $p$ and the links $l$.  As we
are dealing with non-commuting matrices, the product around the square
is meant to be ordered, while because of the trace, the starting point
of this ordering drops out.

To formulate the quantum theory of this system one usually uses the
Feynman path integral.  To construct this, exponentiate the action and
integrate over all dynamical variables
\begin{equation}
Z=\int (dU) e^{-\beta S},
\end{equation}
where the parameter $\beta$ controls the bare coupling.  This converts
the three space dimensional quantum field theory of gluons into a
classical statistical mechanical system in four space-time dimensions.
Such a many-degree-of-freedom statistical system cries out for Monte
Carlo simulation, which now dominates the field of lattice QCD.  Note
the close analogy with a magnetic system; we could think of our
matrices as ``spins'' interacting through a four spin coupling
expressed in terms of the plaquettes.

The formulation is conventionally taken in Euclidean four-dimensional
space.  In effect this replaces the time evolution operator $e^{-iHt}$
by $e^{-Ht}$.  Despite involving the same Hamiltonian $H$, excited
states are inherently suppressed and information on high energy
scattering is particularly hard to extract.  However low energy states
and matrix elements are the natural physical quantities to explore
numerically.  This is the bread and butter of the lattice theorist.
Indeed, the simulations reproduce the qualitative spectrum of stable
hadrons quite well\cite{Durr:2010vn}.  Matrix elements currently under
intense study are playing a crucial role in ongoing tests of the
standard model of particle physics.

\section{A paucity of parameters}

It is important to emphasize one of the most remarkable aspects of
QCD: the small number of adjustable parameters.  To begin with, the
lattice spacing itself is not an observable.  We are using the lattice
to define the theory, and thus for physics we must take the continuum
limit $a\rightarrow 0$.  Then there is the coupling constant, which is
also not a physical parameter due to the phenomenon of asymptotic
freedom.  The lattice works directly with a bare coupling, and in the
continuum limit this should vanish as predicted by asymptotic freedom
\begin{equation}
g_0^2 \sim {1\over \log(1/\Lambda a)} \rightarrow 0.
\end{equation} 
In the process, the coupling is replaced by an overall scale
$\Lambda$, which can be regarded as an integration constant for the
renormalization group equation.  Coleman and
Weinberg\cite{Coleman:1973jx} gave this phenomenon of replacing a
dimensionless coupling with a scale the marvelous name ``dimensional
transmutation.''  An overall scale is not really something we should
expect to calculate from first principles.  Its numerical value would
depend on the units chosen, be they furlongs or light-fortnights.
 
Next consider the quark masses.  These also renormalize to zero as a
power of the coupling in the continuum limit.  Factoring out this
divergence, we can define a renormalized quark mass, a second
integration constant of the renormalization group equations.  One such
constant $M_i$ is needed for each quark ``flavor'' or species $i$.  Up
to an irrelevant overall scale, the physical theory is then a function
only of the dimensionless ratios $M_i/\Lambda$.  These are the only
free parameters in the strong interactions.  The origin of the
underlying masses remains one of the outstanding mysteries of particle
physics.

With multiple flavors, the massless quark limit gives a rather
remarkable theory, one with no undetermined dimensionless parameters.
This limit is not terribly far from reality; chiral symmetry breaking
should give massless pions, and experimentally the pions are
considerably lighter than the next hadron, the rho.  A
theory of two massless quarks is a fair approximation to the strong
interactions at intermediate energies.  In this limit all
dimensionless ratios should be calculable from first principles,
including quantities such as the rho to nucleon mass ratio.

Since it is absorbed into an overall scale, the strong coupling
constant at any physical scale is not an input parameter, but should
be determined from first principles.  Such calculations have placed
lattice gauge theory into the famous particle data group
tables\cite{Agashe:2014kda}.

\section{Numerical simulation}

While other techniques exist, such as strong coupling expansions,
large scale numerical simulations currently dominate the practice of
lattice gauge theory.  They are based on evaluating the path integral
\begin{equation}
Z=\int (dU)\ e^{-\beta S}
\end{equation}
with $\beta$ proportional to the inverse bare coupling squared.  A
direct evaluation of such an integral has pitfalls.  At first sight,
the basic size of the calculation is overwhelming.  Considering a
$10^4$ lattice, small by today's standards, there are 40,000 links.
On each is an $SU(3)$ matrix, parameterized by 8 numbers.  Thus we
have a $10^4\times 4 \times 8 = 320,000$ dimensional integral.  One
might try to replace this with a discrete sum over values of the
integrand.  If we make the extreme approximation of using only two
points per dimension, this gives a sum with
\begin{equation}
2^{320,000}=3.8\times 10^{96,329}
\end{equation}
terms!  Of course, computers are getting pretty fast, but one should
remember that the age of universe is only $\sim 10^{27}$ nanoseconds.

These huge numbers suggest a statistical treatment.  The above
integral is formally a partition function.  Consider a more familiar
statistical system, such as a glass of beer.  There are a huge number
of ways of arranging the atoms of carbon, hydrogen, oxygen, etc.~that
still leave us with a glass of beer.  We don't need to know all those
arrangements, we only need a dozen or so ``typical'' glasses to know
all the important properties.

This is the basis of the Monte Carlo approach.  The analogy with a
partition function and the role of ${1\over \beta}$ as a temperature
enables the use of standard techniques to obtain ``typical''
equilibrium configurations, where the probability of any given
configuration is given by the Boltzmann weight
\begin{equation}
P(C)\sim e^{-\beta S(C)}.
\end{equation}
For this we use a Markov process, making changes in the current
configuration
\begin{equation}
C\rightarrow C^\prime \rightarrow \ldots
\end{equation}
biased by the desired weight.

The idea is easily demonstrated with the example of $Z_2$ lattice
gauge theory.\cite{Creutz:1979kf}  For this toy model, the links
are allowed to take only two values, either plus or minus unity.
One sets up a loop over the lattice variables.  When looking at a
particular link, calculate the probability for it to have value $1$
\begin{equation}
P(1)={e^{-\beta S(1)}\over e^{-\beta S(1)}+e^{-\beta S(-1)}}.
\end{equation} 
Then pull out a roulette wheel and select either 1 or $-1$ biased by
this weight.  Lattice gauge Monte-Carlo programs are by nature quite
simple.  They are basically a set of nested loops surrounding a random
change of the fundamental variables.

Extending this to fields in larger manifolds, such as the $SU(3)$
matrices representing the gluon fields, is straightforward.  The
algorithms are usually based on a detailed balance condition for a
local change of fields taking configuration $C$ to configuration
$C^\prime$.  If probabilities for making these changes in one step satisfy
\begin{equation}
{P(C\rightarrow C^\prime)\over P(C^\prime\rightarrow C)}
={e^{-\beta S(C^\prime)}\over e^{-\beta S(C)}},
\end{equation}
a simple argument shows that under this condition any ensemble of
configurations will approach the equilibrium ensemble.

The results of these simulations have been spectacular, giving first
principles calculations for interacting quantum field theories.  I
will just mention a few examples.  The early result that bolstered the
lattice into mainstream particle physics was the convincing
demonstration of the confinement phenomenon.  The force between two
quark sources indeed remains constant at large distances.

A major goal of lattice simulations is to understand the hadronic
spectrum.  This is done by studying the long distance behavior of
correlation functions.  Let $\phi(t)$ be an operator that can create
a specific particle at time $t$.  Then as $t$ becomes large the
correlator
\begin{equation}
\langle \phi(t)\phi(0)\rangle \propto e^{-mt}
\end{equation}
where $m$ is the mass of the lightest hadron that can be created by
$\phi$.  In these calculations the bare quark masses are parameters
that can be determined by fitting a few of the light mesons.  Chiral
symmetry is useful here, with the pion mass squared predicted to be
proportional to the light quark mass.  Using the pion mass to fix the
light quark mass and the kaon mass to fix the strange quark, all other
particle masses should be determined.  In this way, recent simulations
with physical mass pions have successfully mapped out much of the low
energy hadron spectrum. \cite{Durr:2010vn}

Another accomplishment for which the lattice excels over all other
methods has been the study (using an approximation to QCD) of the
deconfinement of quarks and gluons into a plasma at a temperature of
about 170--190 MeV\cite{Aoki:2006br}.  Indeed, the lattice is the
unique quantitative tool capable of making precise predictions for the
value this temperature.  The method is based on the fact that the
Euclidean path integral in a finite temporal box directly gives the
physical finite-temperature partition function, where the size of the
box is proportional to the inverse temperature.  This transition
represents the confining flux tubes becoming lost in a background
plasma of virtual flux lines.

\section{Concluding remarks}

In summary, lattice gauge theory provides the dominant framework for
investigating non-perturbative phenomena in quantum field theory.  The
approach is currently dominated by numerical simulations, although the
basic framework is more general.  With the recent developments towards
implementing chiral symmetry on the lattice, including domain-wall
fermions, the overlap formula, and variants on the Ginsparg-Wilson
relation, parity conserving theories, such as the strong interactions,
are fundamentally in quite good shape.

A particularly fascinating unsolved issue is the chiral gauge problem.
Without a proper lattice formulation of a chiral gauge theory, it is
unclear whether such models make any sense as a fundamental field
theories.  This is important for understanding how neutrinos can
couple in only one helicity state.  A marvelous goal would be a fully
finite, gauge invariant, and local lattice formulation of the standard
model.  The problems encountered with chiral gauge theory are closely
related to similar issues with super-symmetry, another area that does
not naturally fit on the lattice.  Understanding these issues will be
necessary to make ties with the explosive activity in string theory
and a possible regularization of gravity.

The other major unsolved problems in lattice gauge theory are
algorithmic.  Current fermion algorithms are extremely awkward and
computer intensive.  It is unclear why this has to be so, and may only
be a consequence of our working directly with fermion determinants.
One could to this for bosons too, but that would clearly be terribly
inefficient.  At present, the fermion problem seems completely
intractable when the fermion determinant is not positive.  This is of
more than academic interest since interesting superconducting phases
are predicted at high quark density.  Similar sign problems appear in
other fields of physics, such as doped strongly coupled electron
systems.

Finally, throughout history the question of ``what is elementary?''
continues to arise.  This is almost certainly an ill posed question,
with variant approaches being simpler in distinct contexts.  At a more
mundane level, for low energy chiral dynamics we lose nothing by
considering the pion as an elementary pseudo-goldstone field, while at
extremely short distances string structures may become more
fundamental.  Quarks and their confinement may only be a useful
temporary construct along the way.

\blockcomment
\section*{Acknowledgment}{This manuscript has been authored under
contract number DE-AC02-98CH10886 with the U.S.~Department of Energy.
Accordingly, the U.S. Government retains a non-exclusive, royalty-free
license to publish or reproduce the published form of this
contribution, or allow others to do so, for U.S.~Government purposes.}
\endcomment

\bibliographystyle{ws-procs975x65}

\end{document}